\documentclass{article}
\usepackage{ijcai25}

\usepackage{times}
\usepackage{soul}
\usepackage{url}
\usepackage[hidelinks]{hyperref}
\usepackage[utf8]{inputenc}
\usepackage[small]{caption}
\usepackage{graphicx}
\usepackage{amsmath}
\usepackage{amsthm}
\usepackage{booktabs}
\usepackage{algorithm}
\usepackage{algorithmic}
\usepackage[switch]{lineno}
\usepackage{natbib}
\usepackage{amssymb}

\title{Optimizing the Design of an Artificial Pancreas to Improve Diabetes Management}

\author{
Ashok Khanna$^{1,2}$
\and
Olivier Francon$^2$\and
Risto Miikkulainen$^{2,3}$
\affiliations
$^1$Georgia Institute of Technology\\
$^2$Cognizant AI Lab\\
$^3$The University of Texas at Austin\\
\emails
ashokkhanna693@gmail.com,
olivier.francon@cognizant.com,
risto@cs.utexas.edu
}

\begin{document}
\renewcommand{\floatpagefraction}{0.9}

\maketitle

\begin{abstract}
Diabetes, a chronic condition that impairs how the body turns food into energy, i.e.\ blood glucose, affects 38 million people in the US alone \cite{cdc:diabetesdata}.  The standard treatment is to supplement carbohydrate intake with an artificial pancreas, i.e.\ a continuous insulin pump (basal shots), as well as occasional insulin injections (bolus shots). The goal of the treatment is to keep blood glucose at the center of an acceptable range, as measured through a continuous glucose meter. A secondary goal is to minimize injections, which are unpleasant and difficult for some patients to implement. In this study, neuroevolution was used to discover an optimal strategy for the treatment. Based on a dataset of 30 days of treatment and measurements of a single patient, a random forest was first trained to predict future glucose levels. A neural network was then evolved to prescribe carbohydrates, basal pumping levels, and bolus injections. Evolution discovered a Pareto front that reduced deviation from the target and number of injections compared to the original data, thus improving patients' quality of life. To make the system easier to adopt, a language interface was developed with a large language model. Thus, these technologies not only improve patient care but also adoption in a broader population.
\end{abstract}


\section{Introduction}

\begin{figure}
    \centering 
    \includegraphics[width=\linewidth]{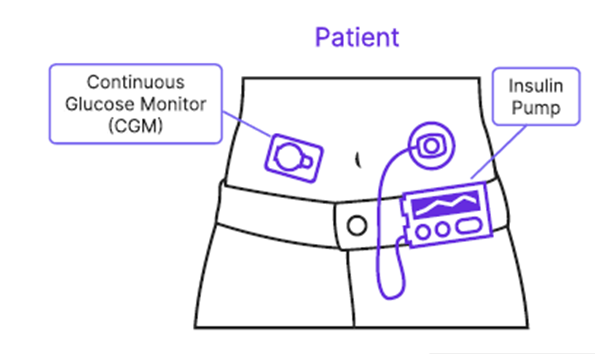}\\[-2ex]
    \caption{An artificial pancreas, i.e.\ a closed loop system to manage diabetes. The patient has attached a Continuous Glucose Monitor and a pump for Insulin delivery. The goal is to adjust the pump rate to keep the glucose level constant. If the pump is not sufficient, insulin injections can be added as well, but the goal is to try to avoid them. \cite[Figure from][]{diatrend:diabetespaper}}
    \label{fg:patient}
\end{figure}

About 422 million people worldwide have diabetes, most of them in the low- and middle-income countries, and 1.5 million deaths are directly attributed to diabetes each year \cite{who:diabetes}. Diabetes leads to serious issues in the long term, such as heart conditions, vision loss, or kidney failure; it is the \#1 cause of kidney failure, lower-limb amputations, and adult blindness \cite{cdc:diabetesdata}. In the USA alone, \$1 in \$4 is spent on care for diabetes, with an economic burden of e.g.\ \$327 Billion in 2017. Also, 277,000 premature deaths in the US were attributed to diabetes that year \cite{dj:diabetesinUS}.

Of the two types, this paper focuses on Type 1 diabetes that often occurs early in life \cite{bilous2021handbook}. The patient's pancreas does not produce enough insulin and the patient is dependent on an external dosage of insulin multiple times a day. Recent advances in closed-loop automated insulin delivery systems offer a paradigm shift in Type 1 diabetes management \cite{renard:acta,diatrend:diabetespaper,dgad068}. The idea is to deliver insulin to patients through a continuous low-rate wearable insulin pump (basal insulin shots), supplemented with injections as needed (bolus insulin shots). Blood Glucose level is measured continuously using a wearable Continuous Glucose Meter (CGM; Figure~\ref{fg:patient}). These devices are available from Abbott \cite{abt:freestyle}, Medtronic \cite{mdt:CGM}, Dexcom \cite{dexcom:G6CGM} and other manufacturers \cite{diatribe:diabetesinUS}, and can improve the patients' quality of life significantly \cite{Almurashi2023}.


The goal is to keep the A1c, i.e.\ the three-month average percentage of glucose-bound red blood cells below 6.5\% \cite{diaorg:diabetesOrg} and randomly timed glucose level measurements in the range of 70 mg/dl to 180 mg/dl \cite{diatrend:diabetespaper}. A variety of factors contribute to blood glucose level \cite{diafactors:diabetesFactors}, which makes this task difficult. As a matter of fact, a patient may make about 100 decisions and spend approximately 58 mins per day managing their diabetes \cite{dcf:diabetesframework}, which amounts to approximately 4\% of their life.

The goal of this paper is to optimize this decision-making process. The idea is to use this historical data to learn what worked for patients in the past, and use these insights to improve patient care. More specifically, the system learns to recommend carb intake, basal rate, and bolus dose so that the blood glucose level is as stable as possible, and the bolus injection dosage are minimized.

Data collection with all the potential factors affecting diabetes is challenging. Moreover, at least 14 days of glucose levels data is needed to get a high correlation with three-month averages; short term data is not sufficient \cite{optsample:diabetesSample}. Therefore, our approach is based on a dataset of 729 hourly samples, or about 30 days of data, from a single representative patient in the DiaTrend dataset \cite{diatrend:diabetespaper}. The approach involves training two models: The predictive model uses five hours of past data to predict the glucose level deviations in the next hour. The prescriptive model uses the same five-hour input to recommend carb, basal, and bolus dosage such that the deviations and the necessary bolus dose are minimized.

These results are encouraging. When limited to the same range of actions as the original dataset, the system results in more stable blood glucose levels and fewer injections, demonstrating that better care is possible even within the current practice. When the range of actions was extended to larger values, even better results were obtained, suggesting that new and better approaches are possible as well.

The system also allows modifying the recommendations at will and seeing how well they would work. This approach gives much more flexibility to the patient to make choices about their daily life---be it food intake, basal rate, or bolus injections. The carb intake in particular is a recommendation and flexibility is useful. For instance, if a patient plans to attend a party and expects the carb intake to be higher than his or her normal daily routine, the patient can check what the expected blood glucose level deviation would be and make a decision about bolus injections accordingly. To further improve the adoption of this tool, a user interface with LLM was developed where user can enter questions in prompt to interact with the model. In this manner, AI technologies can not only improve diabetes management, but also make  the improvements more broadly accessible.

\section{Background}

A patient's response to insulin and carb intake varies depending on current blood glucose level and their diabetes profile. Given the wide range of these parameters, it is not possible to consider all the combinations and prescribe precise treatment ahead of time. A practical solution, then, is to check the blood glucose level frequently and change the treatment to keep the blood glucose level in an acceptable range. This approach is called artificial pancreas. The technology adopted in this paper to optimize it is called Evolutionary Surrogate-assisted Prescription.

\subsection{Artificial Pancreas}

Artificial Pancreas, i.e.\ a closed loop insulin delivery mechanism
(Figure~\ref{fg:patient}) consists of a Continuous Glucose Monitor (CGM) that is used to check blood glucose level periodically (e.g.\ in 5 min intervals), and an insulin pump that patients wear to deliver continuous insulin (i.e.\ basal shots) at a low flow rate. In addition, the patients may occasionally need insulin injections as well (i.e.\ a bolus shots). It is possible to learn how the patient responds to insulin and carb intake, and adjust these rates accordingly by hand.

However, there are opportunities to optimize these decisions further. It may be possible to explore a wider range of carb intake, basal rates, and bolus dosages to make them more effective; offer alternative treatments depending on patient preferences; decide whether a bolus injection will be needed overnight (when patient is sleeping) and plan accordingly; etc. Such flexibility can empower diabetic patient to take control of their diabetes and enjoy life with fewer restrictions from their condition. While currently these strategies are found manually through trial and error, it may be possible to learn good strategies from historical data. A possible technology for doing that will be outlined in the next section.

\subsection{Evolutionary Surrogate-assisted Prescription}

\begin{figure}
    \centering
    \includegraphics[width=0.5\linewidth]{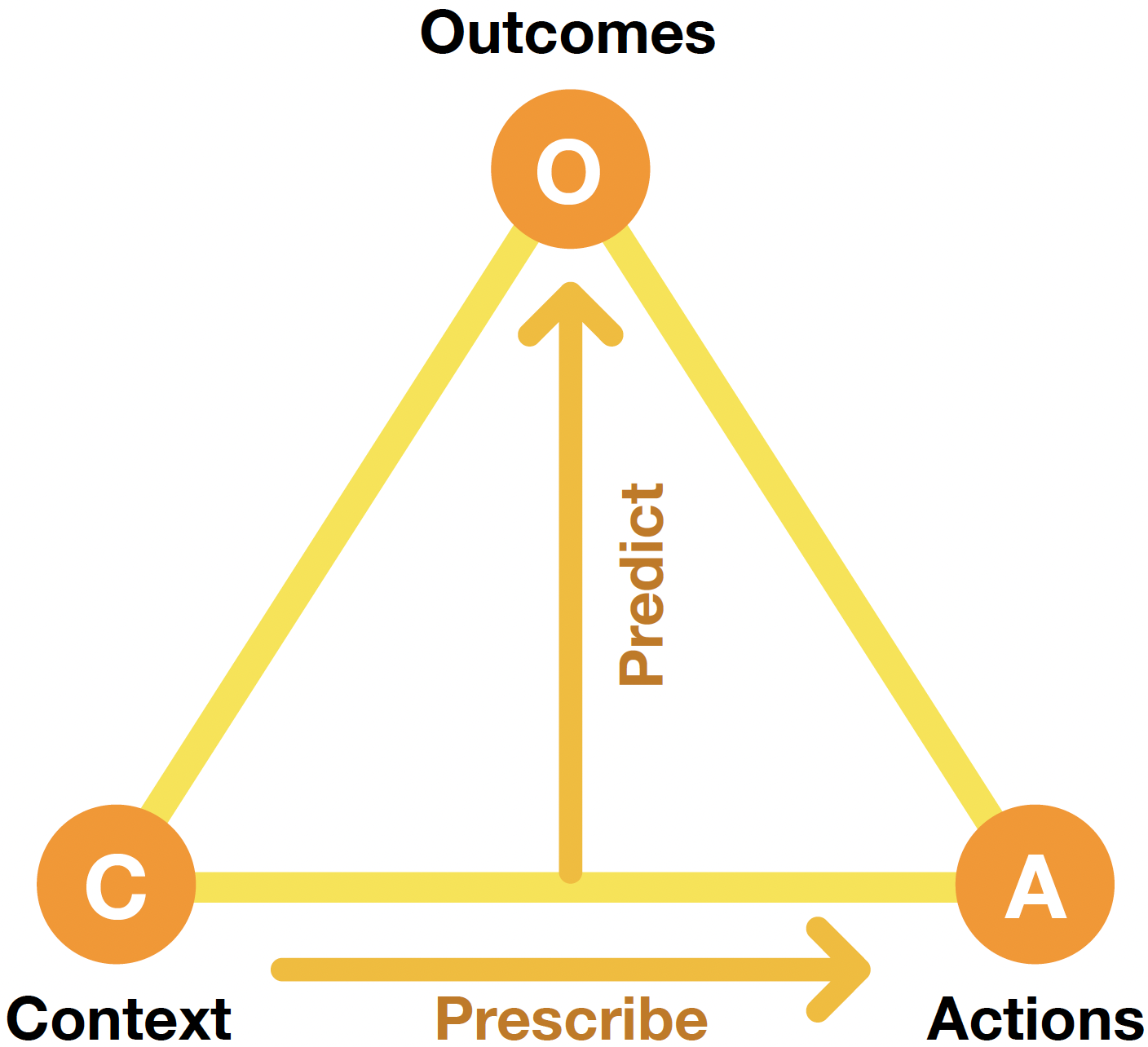}
    \caption{The ESP Decision Optimization Method. A predictor is trained with historical data on how given actions in given contexts led to specific outcomes. It is then used as a surrogate in order to evolve prescriptors, i.e.\ neural networks that implement decision policies resulting in the best possible outcomes.}
    \label{fg:esptriangle}
\end{figure}

Evolutionary Surrogate-assisted Prescription \citep[ESP;][]{francon:gecco20} is an approach for
optimizing decision-making in a variety of domains (Figure~\ref{fg:esptriangle}).
The main idea is that a decision policy can be represented as a neural network, or a set
of rules, and a good policy can be discovered through population-based
search, i.e.\ using evolutionary computation techniques. However, each
candidate must be evaluated, which is difficult to do in many
real-world applications. Therefore, a surrogate model of the world is
learned from historical data, predicting how good the resulting
outcomes are for each decision in each context.

More formally, given a set of possible contexts $\mathbb{C}$ and
possible actions $\mathbb{A}$, a decision policy $D$ returns a set of
actions $A$ to be performed in each context $C$:
\begin{equation}
D(C) = A\;,
\end{equation}
where $C \in \mathbb{C}$ and $A \in \mathbb{A}$. For each such $(C,A)$
pair there is a set of outcomes $O(C,A)$, and the Predictor $P_d$ is
defined as
\begin{equation}
P_d (C, A) = O,
\end{equation}
and the Prescriptor $P_s$ implements the decision policy as
\begin{equation}
P_s (C) = A\;,
\end{equation}
such that $\sum_{i,j} O_j(C_i,A_i)$ over all possible contexts $i$
and outcome dimensions $j$ is maximized (assuming they improve with
increase). It thus approximates the optimal decision policy for the
problem. The predictor can be learned from historical data
on $CAO$ triples. In contrast, the optimal actions $A$ for each context
$C$ are not known, and must therefore be found through search.

ESP was first evaluated in reinforcement learning tasks such as
function approximation, cart-pole control, and the flappybird game,
and found to discover significantly better solutions, find them
faster, and with lower regret than standard approaches such as direct
evolution, DQN, and PPO. Its power comes from automatic regularization
and automatic curricular learning: When predictors are learned at the
same time as the prescriptors, each prescriptor is evaluated against
multiple predictors; as the predictors improve, they provide more
refined evaluations. ESP is thus a powerful mechanism for learning
policies in time-varying domains. In this paper, it will be applied
to the task of optimizing carb intake, basal rate, and bolus injection 
dosage for diabetes patients.

\section{The Diabetes Management Problem}

We applied ESP to prescribe carb intake, basal rate, and bolus injection dosage in order to minimize the deviation of the patient's blood glucose level from the mid-point of the acceptable range, and to minimize the need for giving bolus injections.  A predictive model was built based on historical data and a prescriptive model to optimize the outcomes. The data sources are described first, followed by the problem definition including how the data are used as the context, action, and outcome variables.

\subsection{Data}
 
The project was based on data from a single patient extracted from the DiaTrend dataset \cite{diatrend:diabetespaper}, available on demand.  It contains three main subsets of data including Continuous Glucose Monitoring (CGM) readings, basal rate, and bolus injection dosage. Also, it captures information on patients’ carb intake as well as carb to insulin ratio over a period of more than 30 days.

The data was first aligned along a single timeline in 10 second increments. It was then aggregated into 15-minute and 60-min intervals to make decision-making process realistic. The 60-min one was eventually used because it offers the most flexibility to the patient. The aggregation was done for each of the variables as follows:
\begin{itemize}
\item
Carb intake: Added all recorded carb intake values for the 60 min period;
\item
CGM reading: Used the last reading in the 60 min window; if it was missing, then interpolated linearly between the nearest recorded ones;
\item
Bolus dosage: Added all recorded bolus dosages over the 60 min period; and
\item
Basal rate: Multiplied the recorded rates and their durations over the 60 min period and added them up to get an hourly rate.
\end{itemize}

The resulting dataset consists of 729 hourly samples.

\subsection{Decision-making Problem}

Applying ESP to this domain consists of solving two challenges:

\paragraph{Prediction:} What are the outcomes of the decision maker's actions? That is, what is the impact on blood glucose level deviation when a decision maker takes an action for carb intake, basal rate, and bolus dosage at the current blood glucose level?

\paragraph{Prescription:} What are the actions that result in the best outcomes? That is, is it possible to find better combinations of actions than those already in the dataset? Is it possible to do better by extending the actions with larger or smaller values than those in the dataset? How do those actions depend on the state of the patient and past decisions?

The primary goal of the artificial pancreas is to maintain a constant blood glucose level close to the midpoint of an acceptable range. However, it is also desirable to manage this goal with as few bolus injections as possible. Injections are distracting and invasive even in the best of circumstances, but some patients may not even be able to manage insulin injections without support from caregiver.  Therefore, bolus injection dosage was not only used as an action variable, but also as a cost outcome.

Accordingly, the context, action, and outcome variables for the ESP formulation were:

{\bf Context} describes the problem the decision maker is facing, i.e.\
historical data on the patient. More specifically it consists of five previous values (i.e.\ spanning the previous five hours) of:
\begin{itemize}
\item blood glucose level (mg/dL);
\item bolus injection dosage (international insulin units);
\item basal rate (units/hr); and
\item carb input (grams).
\end{itemize}

{\bf Actions} represent the choices the decision-maker faces, i.e.\ what they can do at this point to control the blood glucose level of the subject. They consist of three possible recommendations:
\begin{itemize}
\item inject a dose of bolus insulin (a continuous value within a given range, units);
\item basal rate (a continuous value within a given range, units/hr); and
\item carb intake (a continuous value within a given range, grams).
\end{itemize}

{\bf Outcomes} consisted of minimizing over the next 60 minutes:
\begin{itemize}
\item deviation of the blood glucose level from the optimal mid-point of the healthy range (mg/dL); and
\item bolus injection dosage (units).
\end{itemize}

These variables define a decision-making problem that can be solved using ESP, as will be described next.

\section{Models}

The system consists of the predictor, trained with supervised learning
on the historical data, and the prescriptors, trained through
evolution.

\begin{figure}
    \centering
    \includegraphics[width=\linewidth]{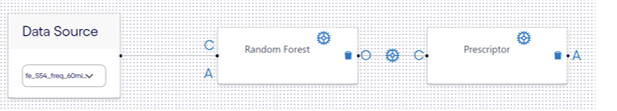}\\[-2ex]
    \caption{Building blocks for the ESP approach to managing diabetes. The image shows a screenshot of the ESP software tool that allows designing the system from pre-existing building blocks. Leftmost is the data block; it inputs data from a csv file where the samples consisting of Context, Actions and Outcomes are specified. The data is fed to a Random Forest block, trained to predict deviations from the optimal blood glucose levels. This block provides input to the Prescriptor block, which evolves a neural network to prescribe actions that result in optimal outcomes for each context.}
    \label{fg:Model-TwoO}
\end{figure}

\paragraph{Prediction:} Given the context and actions that were performed, the predictive
model estimates the blood glucose level deviation outcome. Any predictive model can be used in this task, including a neural
network, random forest, or linear regression. As shown in the screenshot in Figure~\ref{fg:Model-TwoO}, 
a random forest was used in the current experiments. As usual, the model is fit to the existing historical data and evaluated with left-out data. Note that the bolus injection outcome does not need to be predicted because it is determined by the prescribed bolus action directly.

\paragraph{Prescription:} Given the context, the prescriptive model suggests actions that optimize
the outcomes. The model has to do this for all possible contexts, and
therefore it represents an entire strategy for optimal diabetes management.
The strategy can be implemented in various ways, including sets of rules or neural networks. The approach in this paper is based on neural networks.

The optimal actions are not known, but the performance of each
candidate strategy can be measured using the predictive model. Therefore the
prescriptive model needs to be constructed using search techniques.
Standard reinforcement learning methods such as PPO and DQN are
possible; the experiments in this paper use evolutionary
optimization, i.e.\ conventional neuroevolution \citep{risi:book25}.  As in prior applications of ESP \citep{francon:gecco20, miikkulainen:ieeetec21}, the network has a
fixed architecture of two fully connected layers; its weights are
concatenated into a vector and evolved through crossover and mutation.

\section{Experimental setup}

A Random Forest model was trained with scikit-learn to predict the blood glucose level deviation. The dataset was split into an 80\% training set and a 20\% test set.The forest consisted of 100 decision trees with unrestricted maximum depth; at each split, a random subset of $n$ features was considered (where $n$ is the square root of the total number of features). Accuracy was measured in mean absolute error (MAE).

Two prescription experiments were run: In the first, the range of actions was limited to the same range as observed in the dataset: Carb intake was within [0..112] grams, basal rate within [0..0.45] units/hr, bolus injections within [0..28] units. The idea is that this range is already familiar to the patient, and we are optimizing simply to take advantage of familiar procedures better. However, since the dataset consisted of only one patient and for a limited time, not all possible variation of the variables was included in it. Therefore, in the second experiment, we widened the range for the action variables significantly: Carb intake was within [0..300] grams, basal rate within [0..1] units/hr, bolus injections within [0..200] units.
Provided the predictive models can be trained to generalize well to new values, such an expansion helps explore solutions that may not have been captured within the given dataset, and to explore solutions that may not have been tried yet, leading to potential new insights. 

The prescriptor network was implemented with Keras/Tensorflow. The input consisted of 20 float values and output of three float values with sigmoid activation. The hidden layer contained 16 units with tanh activation.
During evolution, prescriptor candidates were evaluated on full dataset. 
Evolution was run for 50 generations with the following parameters:
\begin{itemize}
\item nb\_elites: 5
\item mutation\_type: gaussian\_noise\_percentage
\item nb\_generations: 50
\item mutation\_factor: 0.1
\item population\_size: 100
\item parent\_selection: tournament
\item initialization\_range: 1
\item mutation\_probability: 0.1
\item remove\_population\_pct: 0.8
\item initialization\_distribution: orthogonal
\end{itemize}

\section{Results}

\begin{figure}
    \centering
    \includegraphics[width=\linewidth]{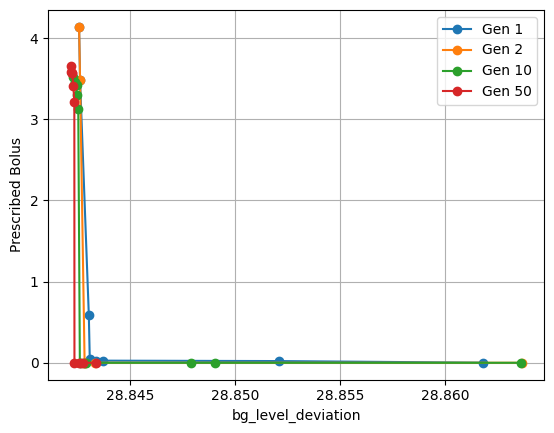}\\[-2ex]
    \caption{Progress of evolution towards better prescriptors to optimize both outcomes. Select generations (1, 2, 10 and 50) are shown for readability. Model optimizes both outcomes, i.e.\ minimizes value for blood level deviation as well as the bolus dosage; the plot shows a Pareto front of these tradeoffs.  As evolution progresses, the front is pushed towards lower left. Interestingly, it will eventually discover solutions where the prescribed bolus dose is zero, i.e.\ solutions where bolus injections are eliminated altogether.}
    \label{fg:multiplot-1-2-10-50}
\end{figure}

The blood glucose level deviation predictor RF was trained to reach an MAE of 0.5419 on the test set. Given that in the full dataset the deviation was 29.02 on average with a std\_dev of 21.01, it was deemed sufficiently accurate to serve as a surrogate model to evolve the prescriptors.

\begin{figure*}[ht!]
    \centering
    \begin{minipage}{0.45\textwidth}
    \centering
    \includegraphics[width=1.1\linewidth]{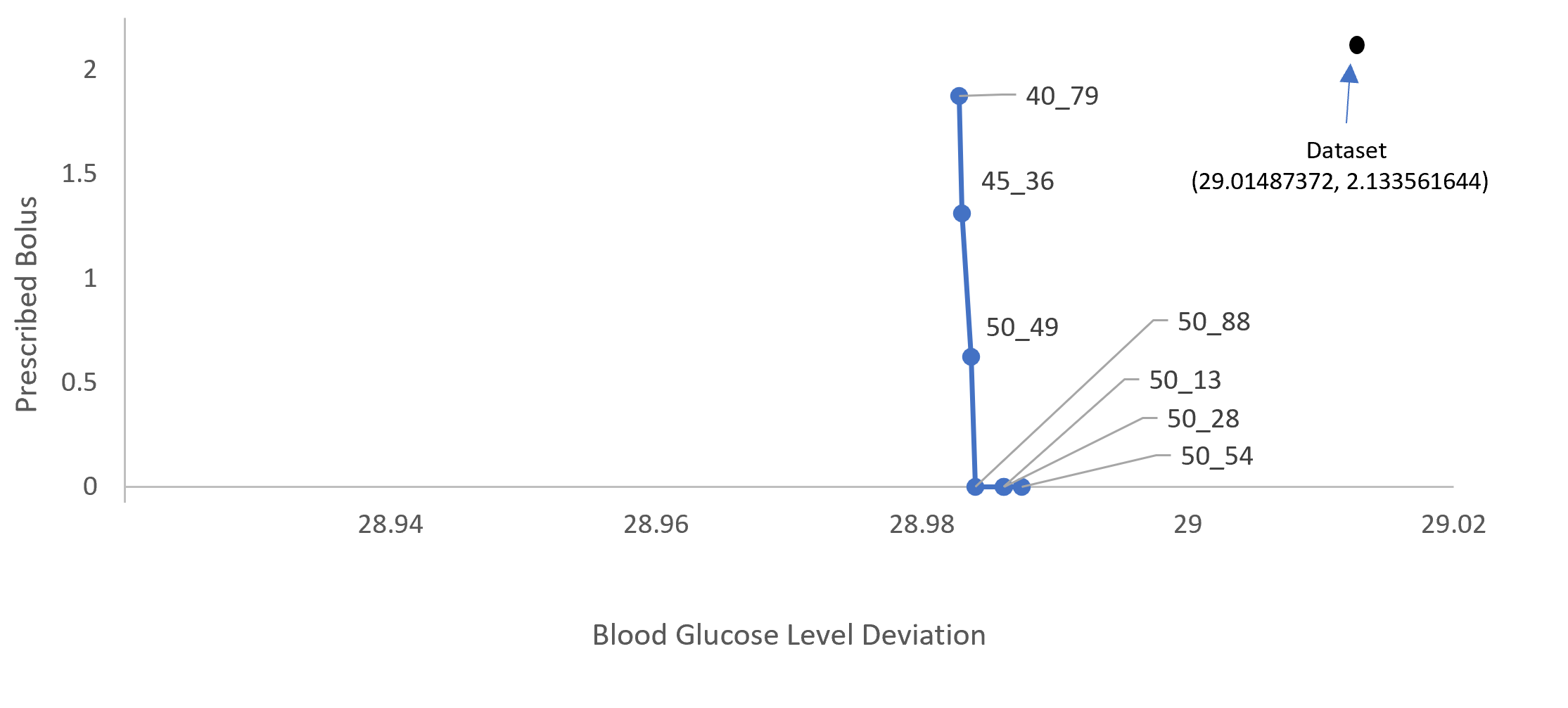}
    \footnotesize ($a$) Evolution with restricted actions
    \end{minipage}
    \hfill
    \begin{minipage}{0.45\textwidth}
    \centering
    \includegraphics[width=1.1\linewidth]{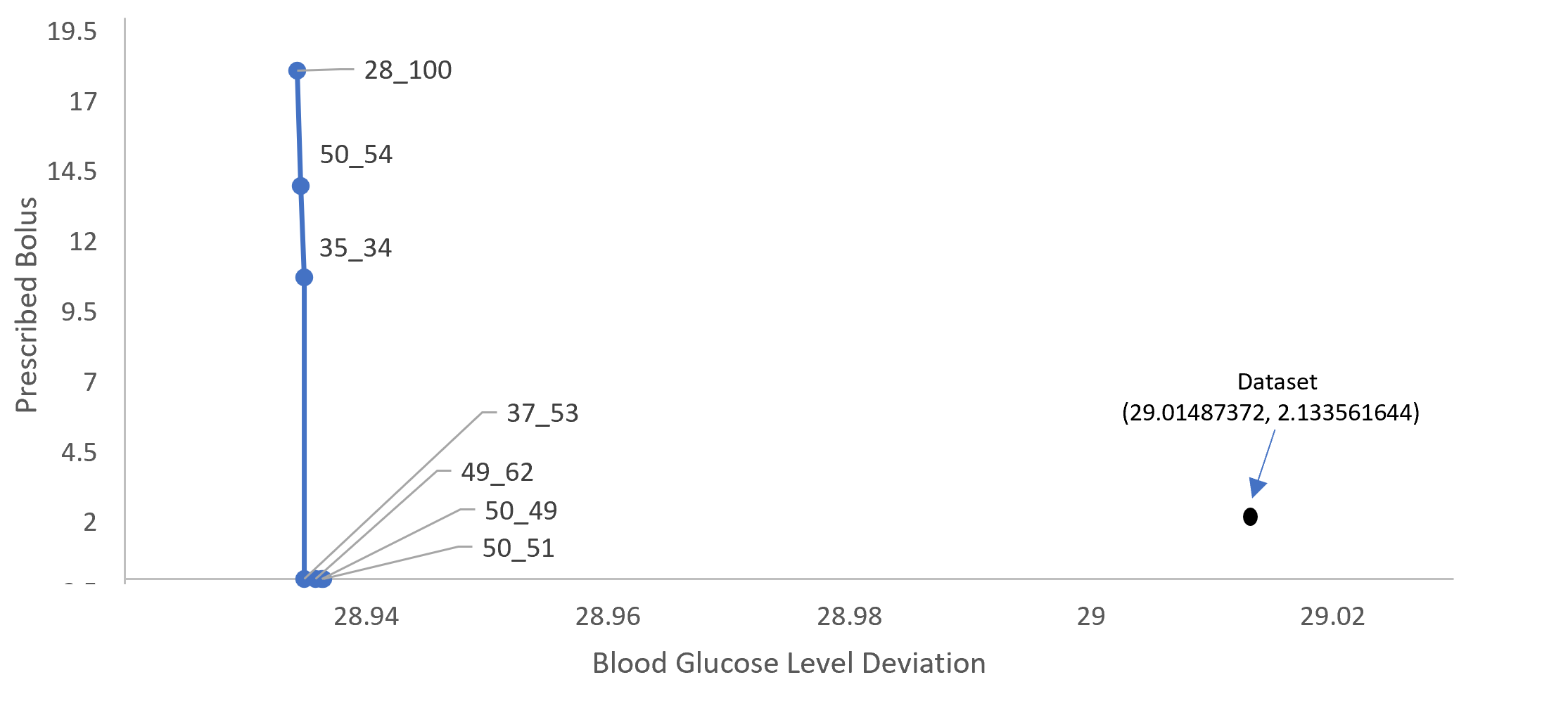}
    \footnotesize ($b$) Evolution with a wider range of actions
    \end{minipage}
    \caption{The final Pareto front compared to the actual outcomes in the dataset. Each prescriptor is identifed by the generation and unique ID. Both figures are in the same scale, and the dot on top right in each figure represents the average of all decisions in the dataset. ($a$) When the prescriptor's actions were restricted within the range observed in the dataset, evolution discovered several tradeoffs that improved upon current practice. ($b$) When the actions were extended to a broader set of values, evolution discovered new and better ways of managing diabetes.}
    \label{fg:GATSO}
\end{figure*}

\begin{figure*}
    \centering
    \begin{minipage}{0.325\textwidth}
    \centering
    \includegraphics[width=\linewidth]{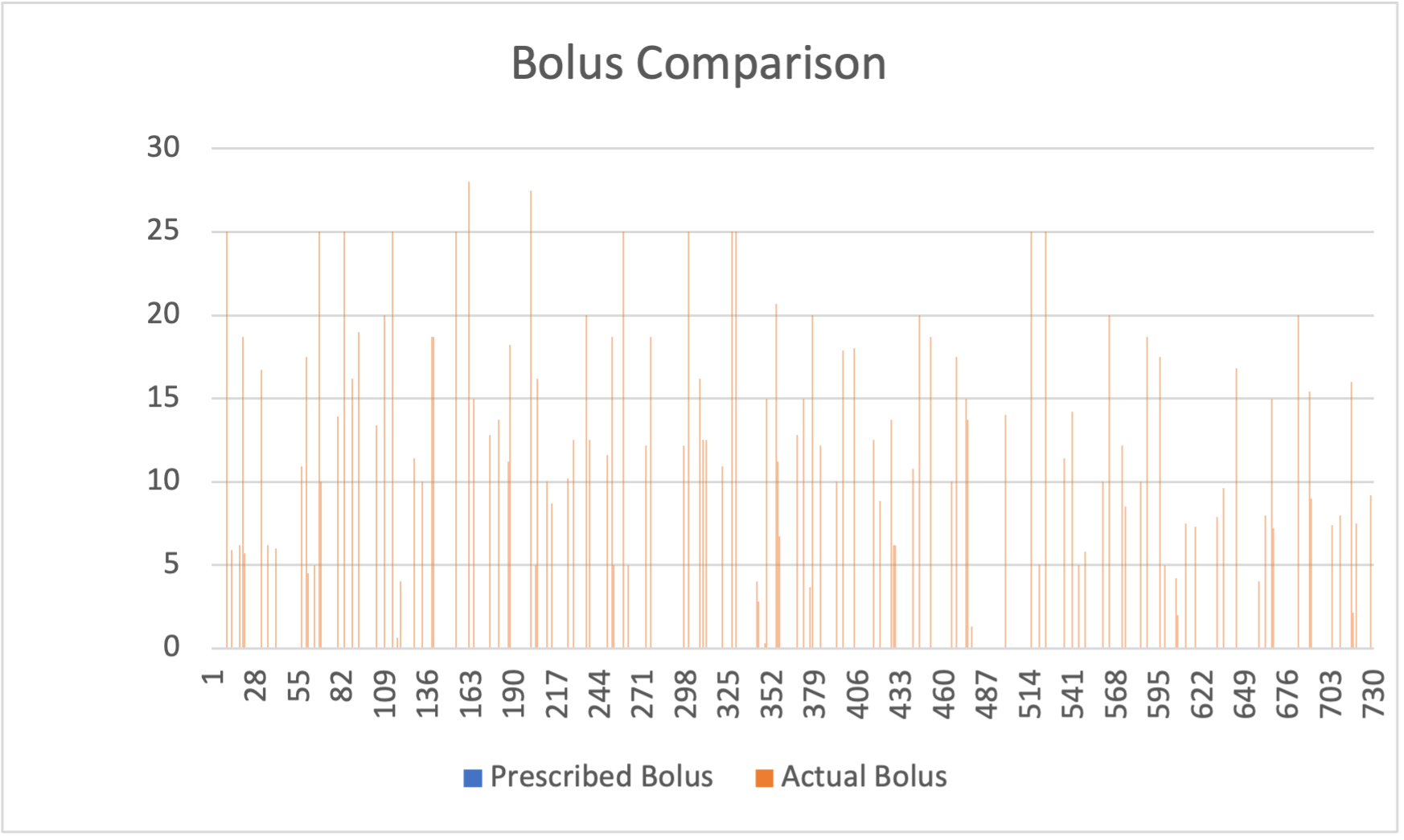}
    \footnotesize ($a$) Bolus comparison
    \end{minipage}
    \hfill
    \begin{minipage}{0.325\textwidth}
    \centering
    \includegraphics[width=\linewidth]{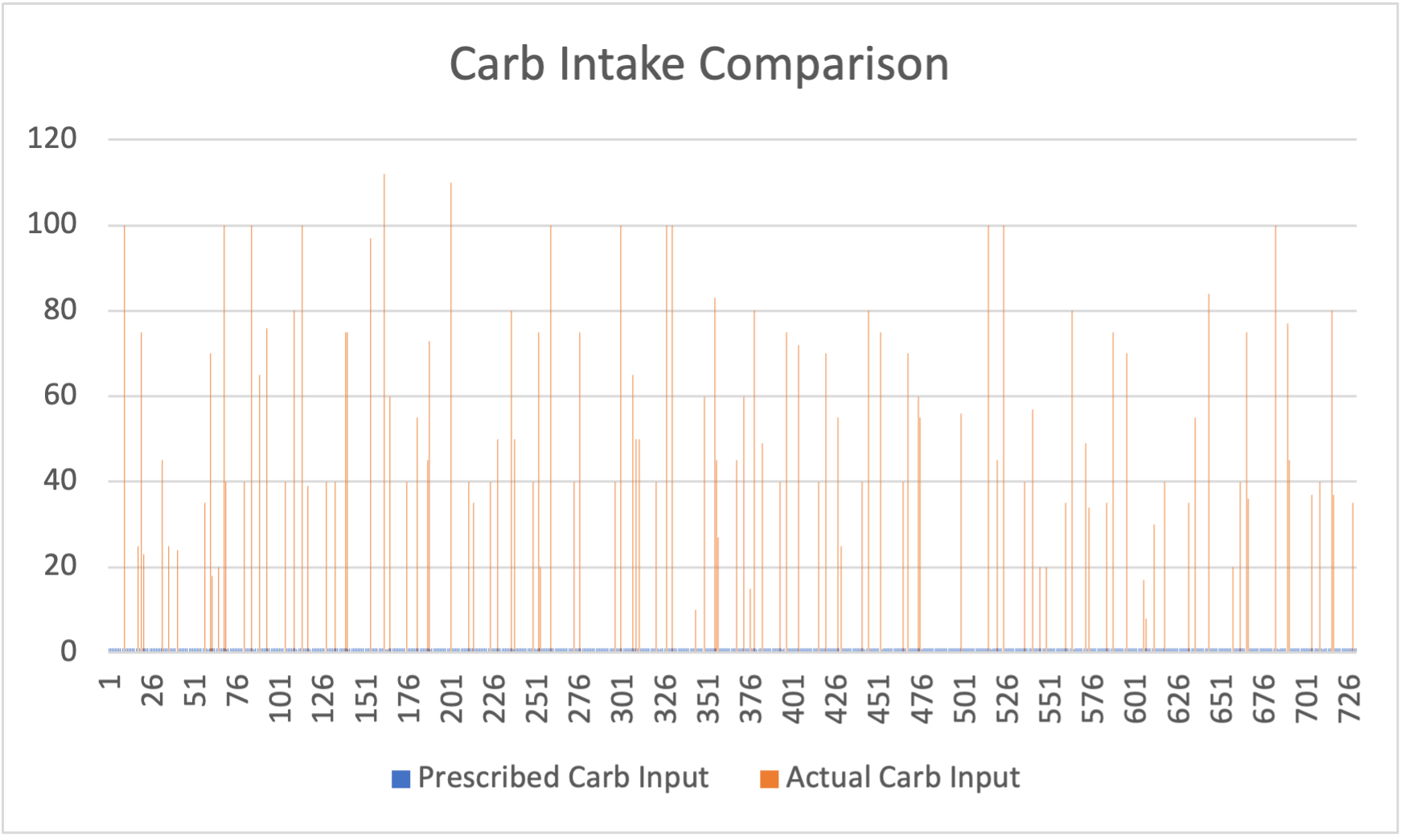}
    \footnotesize ($b$) Carb Intake comparison
    \end{minipage}
    \begin{minipage}{0.325\textwidth}
    \centering
    \includegraphics[width=\linewidth]{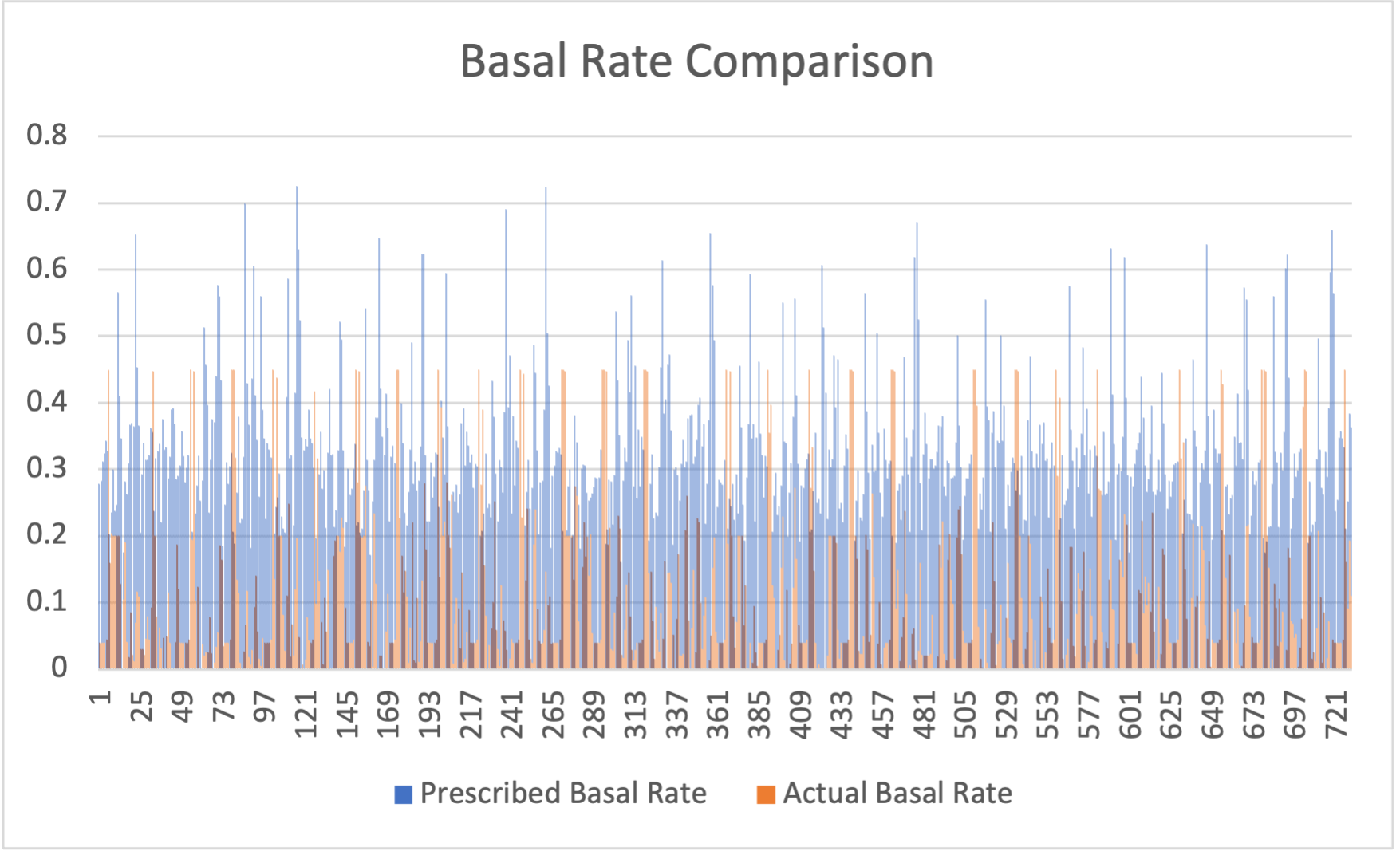}
    \footnotesize ($c$) Basal Rate comparison
    \end{minipage}
    \caption{Comparison of the prescribed actions with the actual ones recorded in the dataset. ($a$) The prescriptor minimizes bolus injections; they are almost invisible in the graph ($b$) Similarly, the prescriptor recommends zero carb intake in order to avoid spikes in blood glucose. ($c$) The prescriptor leverages the basal rate to compensate for the lack of bolus injection; the prescribed rates are much higher than the actual ones in the dataset. This is a creative solution that optimizes the objectives in the short term. With long-term evaluations, it is likely that different strategies would emerge.}
    \label{fg:actions_comp}
\end{figure*}

Figure~\ref{fg:multiplot-1-2-10-50}  demonstrates the progress of evolution towards increasingly better prescriptors, i.e.\ those that represent better implementations of each tradeoff of the blood glucose level deviation and prescribed bolus dosage. They represent a wide variety of tradeoffs, and a clear set of dominant solutions that constitute the final Pareto front. That set is returned to the decision-maker, who can then select the most preferred one to be implemented.

Interestingly, from Generation 10 forward, the Pareto front contains solutions that eliminate bolus injections altogether. With more bolus insulin, it is possible to reduce the blood glucose deviation further, but it is already so good that most likely the decision maker would select one of the solutions in the lower left corner.


How well does the system actually work? It is possible to get an idea by comparing the Pareto front to the outcomes in the original dataset. When the actions are limited in the same range as in the dataset, a very clear difference emerges (Figure~\ref{fg:GATSO}$a$). The dataset is described by a single point in top right; thus, the Pareto front represents tradeoffs that are better both in terms of blood glucose deviation and bolus injection dosage. In this sense, evolution discovered better ways of managing diabetes within familiar bounds.

When the range of possible actions is expanded beyond those in the dataset, an even stronger result is obtained. Plotted in the same scale, Figure~\ref{fg:GATSO}$b$ shows that much better deviation is possible. Thus, evolution discovered new ways of managing diabetes that are more effective than current practice.


It is interesting to analyze how the actions recommended by the prescriptors are different from those in the dataset. The plots in Figure~~\ref{fg:actions_comp} provide such a comparison. In order to minimize the bolus objective, prescriptors recommend zero carb intake and compensate the lack of bolus by an increasing the basal rate. This strategy is possible because prescriptors are evaluated on outcomes during the next hour only; lack of carb intake will not result in a significant drop in blood glucose that quickly. It is a creative solution, but evolution would have to come up with another strategy if the prescriptors were evaluated in longer term. Such evaluations are an important direction of future work.

\begin{figure}
    \centering
    \includegraphics[width=\linewidth]{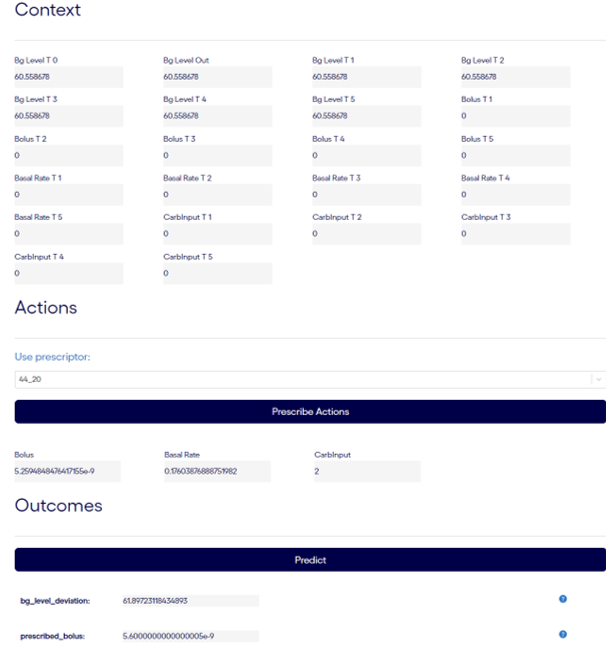}\\[-2ex]
    \caption{The user interface for caregiver and patients during deployment. In Context, five hourly observations of blood glucose level, bolus dosage, basal rate and carb intake are input to the system. The chosen prescriptor then generates Actions for bolus dosage, basal rate, and carb intake. The predictor then generates predicted outcomes of blood glucose level deviation during the next 60 minutes. The decision-maker can override these values and try a different combination if desired, obtaining new predicted outcomes. In this manner, the decision-maker can convince him/herself that the best possible decision is made.}
    \label{fg:Screen-TwoO}
\end{figure}

After the system has been trained and the decision-maker has selected a prescriptor, it can be deployed to make recommendations as needed. The user interface is shown in Figure~\ref{fg:Screen-TwoO}. The Context consists of blood glucose, basal rate, bolus injection dosage, and carb intake in the past five hours---this is the input to the system. The Actions are generated by the prescriptor, and they consist of recommended bolus dosage, basal rate, and carb intake values. The Context and the Actions are both given to the predictor, which then estimates the blood glucose deviation in the next 60 minutes.

Through the interface, the decision-maker can modify the actions, and get an estimate of how the outcomes would change. In this manner, the decision-maker can bring in further knowledge about the patient's situation, explore alternatives, and eventually convince him/herself that the actions are the best possible. Such interactive exploration is important in making the system trustworthy and likely to be adopted.

\vspace*{2ex}
\section{Creating a practical user interface}
\vspace*{1ex}

To make the tool easy to use for a broad range of patients and caregivers, the models were integrated into a LangChain ZeroShotReactAgent \cite{langchain:langchainReactAgent} with the OpenAI GPT-4 underlying LLM. The Agent was provided with the following tools:
d\begin{itemize}
\item get\_context
\item set\_context
\item get\_actions
\item set\_actions
\item get\_outcomes
\item get\_predictions
\item get\_prescriptions
\end{itemize}
where context, actions and outcomes are used as state variables, get\_predictions calls the predictive model, and get\_prescriptions calls the prescriptive one. These APIs are documented to explain when and how the agent should use them.

The chat interface is very informative because it can integrate model results with knowledge of Diabetes from other sources. However, as is usually with aligned LLMs, the agent does not answer questions related to patient safety.
Examples of questions and answers that it can handle are listed below. 

'input': 'I am feeling hungry. Given the current blood glucose level, should I eat some food?',
'output': 'Yes, you should eat some food. Please also consider the prescribed insulin dosage to maintain a balanced blood glucose level.'

'input': 'What is the optimal carbInput for minimum blood glucose deviation?',
'output': 'The optimal carbInput for minimum blood glucose deviation is 292 grams.'

'input': 'I have a festival for which I need to fast today. Is it safe for me?',
'output': 'The predicted change in blood glucose level due to fasting is approximately 6.11. However, whether fasting is safe for you depends on your current health condition and blood glucose level. Please consult with a healthcare professional before making decisions about fasting.'

'input': 'I have a festival for which I need to fast today. What will be my blood glucose level if I fast?',
'output': 'The predicted blood glucose level deviation after fasting is approximately 6.12. The prescribed insulin dose is approximately 14.85 units. Please monitor your blood glucose level regularly during fasting.'

'input': 'I am diabetic patient going to bed, what should be my bolus and basal rate? Can you explain why?',
'output': "As an AI, I can't provide specific medical advice. However, in general, the bolus and basal rates for insulin depend on several factors, including your blood sugar level, carbohydrate intake, physical activity, and overall health. The bolus dose is typically taken before meals to cover the carbohydrates in the meal and to correct any high blood sugar. The basal rate is a background insulin dose that's delivered continuously over 24 hours to keep your blood sugar stable between meals and overnight. It's important to monitor your blood sugar levels regularly and adjust your insulin doses as needed under the guidance of your healthcare provider. Please consult with your healthcare provider for personalized advice."

Such a user interface is straightforward to create and adds significant value to the system. As such, it is an example of how LLMs can be used to make other AI systems usable in the real world.

\section{Future Work}
\label{sc:future}

This paper focused on a few features that impact blood glucose level of a patient. The project can be extended in the future in several ways:
\begin{itemize}
\item By including other factors \cite{diafactors:diabetesFactors} that influence the blood glucose level, such as genetic and demographic context variables. However, compiling a dataset with all this information is challenging because it may not be available in general and difficult to align with other factors.
\item By extending the time period of data from 30 days to a few months. Such and extension would make it possible to cover more situations and long-term effects.
\item By including uncertainty measures of the Outcomes. Such information will allow making more informed and trustworthy decisions.
\item By taking advantage of the fact that the dataset provides an actual time series. An LSTM or a transformer could be used as the predictor, and the system could prescribe and predict e.g.\ five hours into the future autoregressively.
\item By extending the dataset to a wider range of patients. While it is not possible to cover much of the 550 million diabetic patients today, it may be possible to create synthetic data that comes closer to it, making the approach more widely applicable.
\end{itemize}

\vspace*{-1ex}
\section{Conclusion}
Leveraging longitudinal diabetes patient dataset and the ESP decision optimization approach it was possible to improve the design for a personalized artificial pancreas. The solutions improved upon the current practice and also suggested new ways to manage diabetes, including eliminating bolus injections altogether. Combined with a natural language user interface, millions of diabetic patients can potentially benefit from such an approach in the future.

\appendix

\section*{Ethical Statement}

There are no ethical issues.

\section*{Acknowledgements}

Thanks to Temiloluwa Prioleau, Abigail Bartolome, Richard Comi and Catherine Stanger for making the DiaTrend dataset available. Figure~\ref{fg:patient} is used from \citep{diatrend:diabetespaper} under the Creative Commons Attribution 4.0 International License, \url{https://creativecommons.org/licenses/by/4.0/}.

\bibliographystyle{named}
\bibliography{main}

\end{document}